\documentstyle[aps,prb,twocolumn,epsfig]{revtex}

\draft

\begin{document}

\twocolumn[\hsize\textwidth\columnwidth\hsize\csname@twocolumnfalse\endcsname

\title{ Spectroscopic and magnetic mirages of impurities in nanoscopic systems}

\author{K. Hallberg, A. Correa, and C. A. Balseiro}

\address{Instituto Balseiro and Centro At\'omico Bariloche, Comisi\'on
Nacional de Energ\'{\i}a At\'omica, 8400 San Carlos de Bariloche, Argentina.}

\date{\today}
\maketitle

\begin{abstract}
We present exact results for magnetic impurities in nanoscopic systems with focusing
properties. We
  analyze the spectroscopic and magnetic properties of Kondo, intermediate valence and
magnetic
  impurities on a sphere with a metallic surface. Exact calculations show the occurrence of
spectroscopic
  and magnetic mirages at the antipodes of the impurity location. Comparison with 
calculations performed using effective models validates previous calculations of 
spectroscopic mirages.
Our results can be extended to other systems with focusing properties like quantum corrals.  
\end{abstract}

\pacs{PACS numbers: 72.15.Qm, 73.22.-f}
]

Interference of electron wave-functions is a common phenomena in condensed
matter physics which leads to a plethora of effects. Some examples are the
formation of energy bands in crystals, the Friedel oscillations in the
screening of charged impurities or the Aharonov-Bohm effect in mesoscopic
circuits. Up to now, due to technological difficulties, it was not possible
to design and build devises that focus electron wave-functions to
artificially create images in a metal. In a recent and notorious experiment,
Manoharan et al.\cite{Mano} were able to project the image of a Kondo
impurity to a remote location. The devise used in this experiment is an
elliptical quantum corral on a Cu surface. A Co atom on an open Cu surface
behaves as a Kondo impurity and generates a characteristic spectroscopic
signature.\cite{Co,Cu} This signature can be observed by measuring the
tunneling current to a scanning tunneling microscope (STM) tip placed at the
vicinity of the impurity. The signal disappears when the tip is moved away
of the Co impurity a distance of the order of 10 \AA . When a Co atom is
placed at one of the foci of the elliptic corral, the Kondo signature is not
only observed when the STM tip is placed close to the impurity but also when
it is located at the empty focus of the corral indicating the coherent
refocusing of the electronic structure from one focus to the other.

These experiments triggered a number of theoretical works where the
electronic image projection in quantum corrals is analyzed.\cite
{Ag,BW,Esp,PS,alig} All calculations incorporate the Kondo effect in a
phenomenological way, either by including a Kondo resonance at the Fermi
energy or adding a phase shift to the conduction electrons. The quantum
mirages so obtained show that in order to reproduce the spectroscopic
properties at the empty focus, correlations do not play an essential role,
they only provide the signature of the impurity at the Fermi level where the
STM experiment probes the local density of electronic states. The Kondo
nature of the Co impurity is relevant only to generate a resonance at the
Fermi level. A complete many body theory is still lacking and it is of
central importance in order to understand some aspects of the problem. There
are still many open questions particularly concerning the spin dynamics in
these systems: While in the vicinity of the impurity, conduction electron
spins are screening the Co spin, are spins at the empty focus fluctuating
coherently with the impurity spin? Moreover, in the presence of a magnetic
non-Kondo impurity that couples ferromagnetically with conduction electrons
placed at one focus, is there a magnetic mirage at the empty focus?

The aim of this work is to present a many body calculation of impurities in
nanoscopic systems with focusing properties and analyze local density of
states at the Fermi level and spin correlations. The calculation is
performed by exact diagonalization in small clusters using the Lanczos
algorithm\cite{Lancz,dyn}. The elliptic corral is not the only geometry
with focusing properties, other examples are two mutually facing parabolas
at a Cu surface or a thin metallic film covering a rugby ball. In the latter
case, an impurity placed at one tip generates a mirage at the other. A
particular case of this geometry is a sphere with a metallic surface where
any point can be defined as a focus and the mirage is formed at the opposite
point: given a point that defines a pole, all meridians cross at the other
pole being the ``optical '' length of all of them the same. The main
difference between these geometries is the degeneracy of the one electron
states, however a deep understanding of the refocusing properties in one of
them will shed some light into all the others.

We analyze the case of the sphere that is the simplest case and is
appropriate for exact diagonalization. Conduction electrons are confined at
the surface of a sphere of radius $R$, the electron coordinates are the
angles ($\theta ,\varphi $) and a Kondo impurity is placed at $\theta =0$.
The Anderson Hamiltonian reads\cite{And}:

\begin{eqnarray}
H &=&\sum_{l,m,\sigma }\varepsilon _{l}c_{lm\sigma }^{\dagger }c_{lm\sigma
}+\sum_{\sigma }E_{d}d_{\sigma }^{\dagger }d_{\sigma }+Ud_{\uparrow
}^{\dagger }d_{\uparrow }d_{\downarrow }^{\dagger }d_{\downarrow }  \nonumber
\\
&&+\sum_{l,\sigma }V_{l}(c_{l0\sigma }^{\dagger }d_{\sigma }+d_{\sigma
}^{\dagger }c_{l0\sigma })
\end{eqnarray}
where $c_{lm\sigma }^{\dagger }$ creates an electron with quantum numbers $l$%
, $m$,$\sigma $ corresponding to the total angular momentum, the $z$%
-component of the angular momentum and the spin respectively and with energy 
$\varepsilon _{l}=\hbar ^{2}l(l+1)/2mR^{2}$, the operator $d_{\sigma
}^{\dagger }$ creates an electron with spin $\sigma $ at the impurity
orbital with energy $E_{d}$ and Coulomb repulsion $U$. The last term
describes the hybridization with $V_{l}=VY_{l0}(0,0)$ where $Y_{lm}(\theta
,\varphi )$ is the spherical harmonic. For an impurity at the pole ($\theta
=0$), only the $m=0$ states are hybridized with the impurity orbital. In the
summation over $l$ we take a cutoff $l_{\max }$ to avoid including wave
functions oscillating with a characteristic wave length shorter than a
typical interatomic distance $a$, ($(a/R)l_{\max }\simeq 1)$. In what
follows we refer to the impurity site ($\theta =0$) as the occupied focus
and to the point $\theta =\pi $ as the empty focus. Energies are taken in
units of $\hbar ^{2}/2mR^{2}$ and we consider a very large on-site Coulomb
repulsion $U$.

\begin{figure}[ph]
\centerline{\psfig{figure=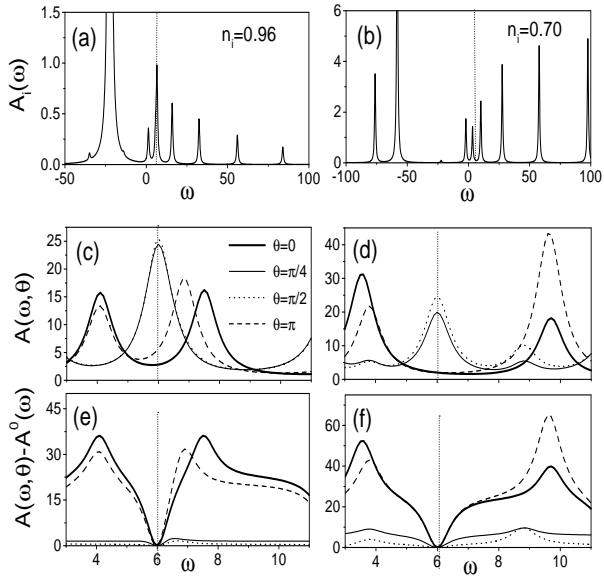,height=10.0cm, width=9cm, angle=0}}
\caption{Spectral density for $E_{d}=-20$ and two values of the
hybridization: $V/4\pi =1$ ((a), (c), (e)) and $V/4\pi =5$ ((b), (d), (f)).
From top to bottom: impurity spectral density; conduction electrons spectral
density for different angles in an expanded scale around $\varepsilon _{F}$
and conduction electrons spectral energy difference with and without
impurity. In (e) and (f) curves have been shifted to coincide at $%
\varepsilon _{F}$. The Fermi energy is indicated with a vertical dotted
line. }
\end{figure}

We first present exact numerical results for the local density of states for
a system with $l_{\max }=10$ and $14$ electrons. In Fig.(1) the impurity
spectral density is shown for different values of the parameters (a
Lorentzian width of 0.5 is taken). A small hybridization $V$ corresponds to
the Kondo regime with an average number of electrons in the impurity orbital 
$<n_{d}>$ close to one. For the parameters of Fig. (1a) we obtain $%
<n_{d}>=0.961$ and the impurity spectral density $A_{i}(\omega )$ shows a
large peak at an energy close to the impurity energy $E_{d}$ and a number of
small peaks at the band energies. Within the band, the dominant peak
coincides with the Fermi energy $\varepsilon _{F}$ and represents the Kondo
resonance. A large hybridization ($V>|E_{d}|$) reduces the number of
electrons a the impurity orbital and for the parameters of Fig (1b) we
obtain $<n_{d}>=0.705$ corresponding to an intermediate valence (IV) regime.
In this case $A_{i}(\omega )$ shows a large amplitude at the band energies
and does not present a dominant Kondo peak. The conduction electron spectral
density $A(\omega ,\theta )$ is shown also in Fig. (1) for different values
of the angle $\theta $. At the impurity focus, the conduction electron
spectral density presents a pseudogap, both for the Kondo and IV regimes
(Figs (1c) and (1d)). As $\theta $ increases the pseudogap is filled and at
the equator ($\theta =\pi /2$) the local density of states resembles the
unperturbed density of states. As $\theta $ approaches $\pi $, the empty
focus, a clear mirage of the impurity is obtained both for small and large $V
$. In this simple geometry, the states that fill the pseudogap for $\theta
\neq 0,\pi $ correspond to the states with $m\neq 0$. For a better
comparison, in Figs. (1e) and (1f) we present the difference $\delta
A(\omega ,\theta )=$ $A(\omega ,\theta )-A^{0}(\omega ,\theta )$ of the
conduction electrons spectral density with and without the impurity. This is
a typical result when the Fermi energy coincides with one of the
non-interacting one-electron levels. When the Fermi energy lies between two
conduction energy levels {\it i.e}. when there is a closed shell in the
sphere, the Kondo peak is not observed and the conduction electron spectral
density shows a small dependence with $\theta $. These results qualitatively
reproduce the experimental observations in the quantum corral where the
mirage is observed only for system sizes such that a confined one-electron
state lies close to the Fremi energy.\cite{Mano}

\begin{figure}[ph]
\centerline{\psfig{figure=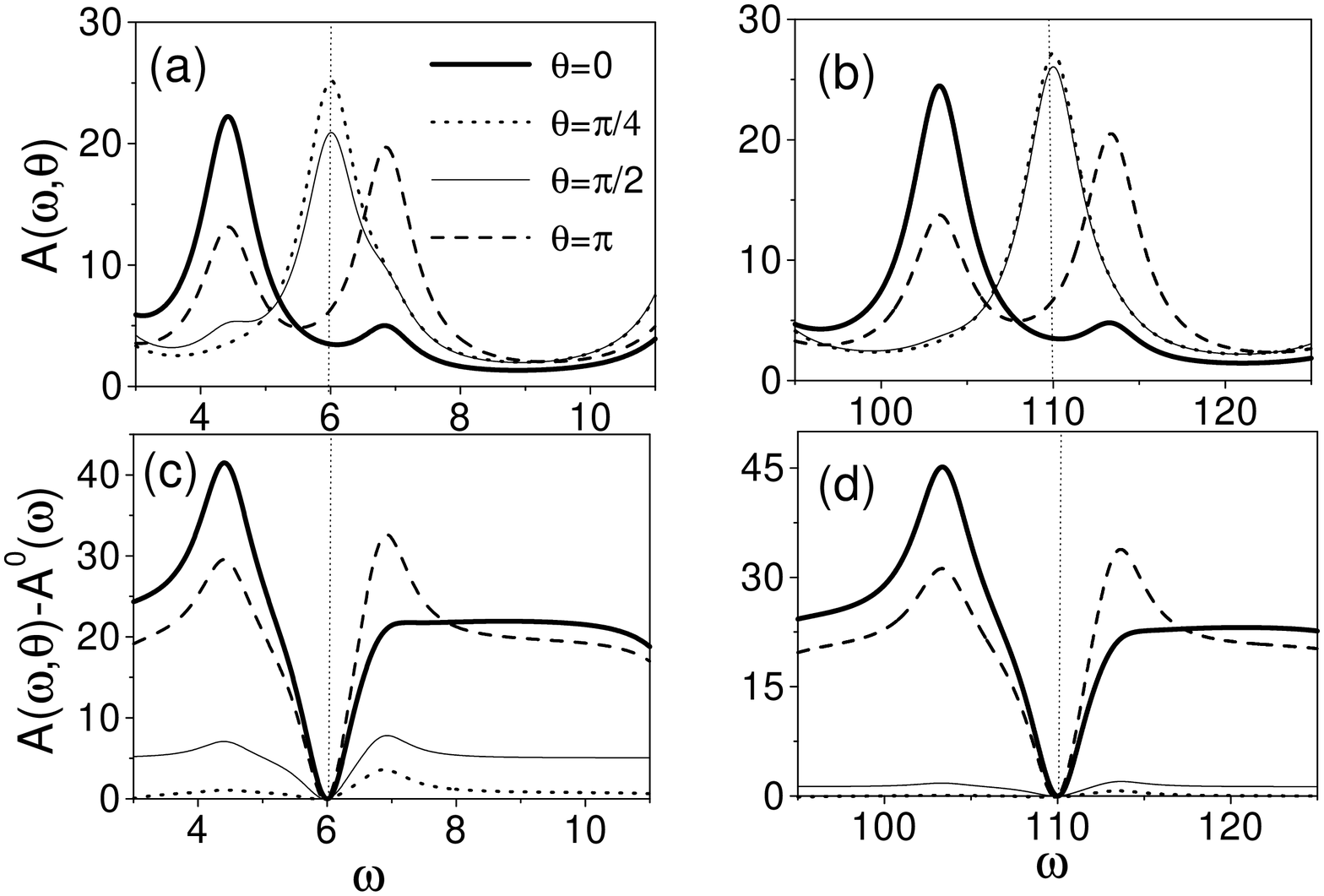,height=6.0cm,width=10cm,angle=0}}
\caption{Conduction electron spectral densities and spectral density
difference for an effective resonant level at $\varepsilon _{F}$. A small
system as in Fig.(1) ((a), (c)) and a large system with $l_{\max }=50$ and $%
220$ electrons ((b), (d)).}
\end{figure}

The low energy spectrum ($\omega \approx \varepsilon _{F}$) obtaind for the
Kondo impurity can be mimicked by representing the Kondo resonance as an
uncorrelated resonant level at the Fermi energy, which is precisely the
spirit of some previous calculations where the impurity propagator is not
fully self-consistently evaluated\cite{BW}. A simple calculation using
Hamiltonian (1) with $E_{d}=\varepsilon _{F}$ , $U=0$ and a small effective
hybridization $V$ gives the results of Fig. (2): for a small system the
spectral density for $\omega \approx \varepsilon _{F}$ qualitatively
reproduce the exact results of Figure (1). Due to the simple structure of
the one electron wave functions on a sphere it is clear that, although the
structure of the difference $\delta A(\omega ,\theta )$ on the two foci for
energies close to the Fermi energy is different, at the Fermi energy $\delta
A(\varepsilon _{F,}0)=\delta A(\varepsilon _{F},\pi )$ independently on the
radius of the sphere provided that the Fermi level coincides with one of the
one-electron levels. An alternative approximate procedure is to express the
conduction-electron propagators in terms of the impurity propagator $%
G_{i}(\omega )$ and then use for $G_{i}(\omega )$ a Lorentzian peak at the
Fermi energy with a width given by the Kondo temperature $T_{K}.$\cite{Ag}
However, in this procedure $G_{i}(\omega )$ does not account for the
structure of the discrete energy levels of the nanoscopic system and the
exact results obtained for an isolated small system are not reproduced. This
approach is justified only if there is a large broadening of the
one-electron levels of the nanoscopic system, a limit in which we expect a
smaller and more diffuse mirage.

These results validate some previous results where the Kondo resonance is
not fully self consistently calculated and the many body effects are taken
into account in a rather phenomenological way. In addition our results show
that also for IV impurities, with a large hybridization or a small $E_{d}$,
the mirage in the spectral density is still obtained.
Another very
interesting feature concerns the spin correlations in these
systems. We start by presenting the static spin correlation $<{\bf S}%
_{imp}\cdot {\bf \sigma }(\theta ,\varphi )>$ where ${\bf S}_{imp}$ is the
impurity spin operator, ${\bf \sigma }(\theta ,\varphi )$ is the conduction
electron spin operator at coordinates $(\theta ,\varphi )$ and the brackets
indicate the expectation value at the ground state. For a small
hybridization the results presented in Figure (3a) clearly show an
antiferromagnetic correlation for small $\theta $ resulting from the
screening of the impurity spin by the conduction electrons. As $\theta $
grows, the spin-spin correlation decreases, however, at the empty focus it
increases again giving rise to a magnetic mirage. At the empty focus, the
conduction electron spin is fluctuating coherently with the impurity spin as
if the Kondo impurity were in its neighborhood. When the Fermi energy
coincides with one of the one electron levels, the Kondo effect takes place
and the conduction electron spin at the empty focus is always
antiferromagnetically correlated with the impurity spin. For the case of a
closed shell, the spin-spin correlations are weaker and at the empty focus
the conduction electron spin is ferromagnetically correlated with the
impurity spin. For a large hybridization charge fluctuations are important
and although the local ($\theta =0$) spin-spin correlations are large the
magnetic mirage is not well defined (dashed line in Fig. (3a)).

\begin{figure}[ph]
\centerline{\psfig{figure=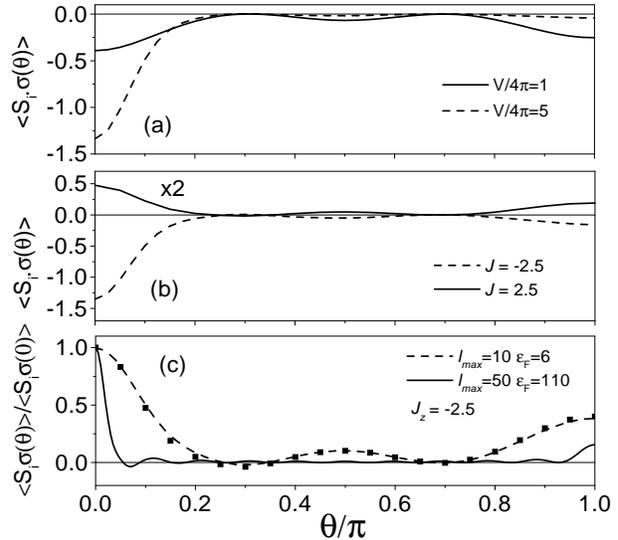,height=8.0cm,width=10cm,angle=0}}
\caption{Spin-spin correlations as a function of $\theta $. (a) Anderson
Hamiltonian (1); (b) Kondo Hamiltonian (2) with ferromagnetic and
antiferromagnetic coupling $J$; (c) Ising-type interaction normalized to the
value at $\theta =0$ for a small (dashed) and large (continuous line)
system. Squares are the exact results obtained in the full Hamiltonian (2).}
\end{figure}

In the Kondo regime the Anderson Hamiltonian can be approximated by the
Kondo Hamiltonian which describes correctly the spin dynamics of the system%
\cite{SW}:

\begin{equation}
H=\sum_{l,m,\sigma }\varepsilon _{l}c_{lm\sigma }^{\dagger }c_{lm\sigma }+J%
{\bf S}_{imp}\cdot {\bf \sigma }(0,0)  \label{Kondo}
\end{equation}
here $J>0$ is the antiferromagnetic coupling between the impurity and the
conduction electron spins. For $J<0$, this Hamiltonian describes the effect
of a magnetic non-Kondo impurity with a ferromagnetic (FM) $sd$-coupling 
\cite{And}. We have also calculated the spin correlations for a magnetic
non-Kondo impurity with FM coupling to the conduction electrons. The results
obtained for $J>0$ and $J<0$ are compared in Fig. (3b). The spin correlation
at short distances (small $\theta $) are larger for the Kondo impurity. This
is due to the singular nature of the Kondo scattering which is not obtained
for a ferromagnetic $sd$-coupling. The magnetic mirage is proportionally
more pronounced for the latter case where perturbations or simple
approximations can be used to estimate the spin-spin correlations. For the $%
J<0$ we have calculated the correlations with an Ising-type interaction $%
J_{z}S_{imp}^{z}\sigma ^{z}(0,0)$ and, as shown in Fig. (3c), the results
quantitatively reproduce the exact results obtained with the full
Hamiltonian (\ref{Kondo}). This allows us to use, for the case $J<0$ , the
Ising-type interaction to estimate the spin correlations in much larger
systems and in more complicated geometries. The spin correlations obtained
for a large sphere are also shown in Fig. (3c). The magnetic mirage is quite
pronounced and despite the fact that the two foci can be even further than
50 interatomic distances, the spin correlation for electrons at the empty
focus are as large as the correlations at distances of the order of one or
two interatomic distances from the occupied focus.

A systematic analysis of the behavior of the magnetic mirage can be made by
changing the sphere radius $R$ and keeping the electron density constant. As
the radius increases, the characteristic energy difference $\delta E$
between two consecutive one-electron levels decreases and for some
particular values of $R$ the Fermi level coincides with one of the
one-electron states. We analyzed these cases in which there is an open
(non-filled) shell in the sphere. The results obtained for different radii
are presented in Fig. (4), the inset shows the amplitude of the spin-spin
correlation between the impurity spin and the conduction electron spin at
the empty focus for fixed density (same as in Fig. 1) and $J_{z}=0.25$ as a
function of $R$. At short distances ($\theta \approx 0$) the spin-spin
correlation is sensitive to both, the system size and the cutoff $l_{\max }$
and as $\theta $ increases it becomes independent of $l_{\max }$. At the
empty focus two different regimes can be observed: For small systems, the
level spacing $\delta E$ is smaller than $J_{z}$ and the impurity polarizes
the electron at the Fermi level; the magnetization profile is essentially
given by the one-electron wave function $|Y_{l_{F},m=0}(\theta )|^{2}$ where 
$l_{F}$ is the Fermi angular momentum. Since these wave functions at $\theta
=\pi $ increase as $l_{F}$ increases, the magnetic mirage increases with $R$
for small systems. For large systems, $\delta E<<J_{z}$, the impurity
produces a mixing of states with different $l$ that interfere destructively
and the magnetic mirage tends to cero as $R$ goes to infinity.

\begin{figure}[ph]
\centerline{\psfig{figure=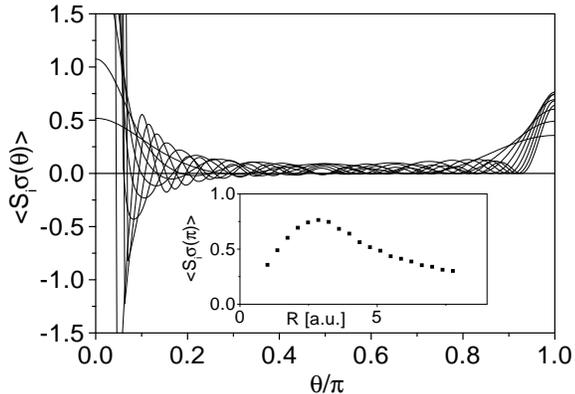,height=6.0cm,width=8cm,angle=0}}
\caption{Spin-spin correlations vs $\theta$ for the Ising limit for
different R with fixed $J_z$ and electron density (open shell). Inset: Spin
correlation at $\theta=\pi$ as a function of R for the same $J_z$}
\end{figure}

As a consequence of the presence of the magnetic mirage, we can infer that
if an additional impurity is placed at the empty focus (i.e. when there are
two impurities in the system, one at each focus), they will strongly
interact. In this case the impurity-impurity coupling is ferromagnetic if
the Fermi level lies close to one of the energy levels of the nanoscopic
system and is antiferromagnetic and much weaker if the Fermi energy lies
between two one-electron levels. Results obtained for the magnetic mirage in
the elliptic quantum corral will be presented elsewhere.

In summary, we have presented here an exact many body calculation for
impurities in nanoscopic systems with focusing properties.  The
spectroscopic characteristics of the Kondo impurity, which are always
observed in its neighborhood for open shell systems, are reproduced at the
opposite point on the sphere giving rise to a spectroscopic mirage of the
impurity.  Our results also show that impurities with a large hybridization,
which induces large charge fluctuations, also give rise to spectroscopic
mirages. This may be important since ab-initio calculations of Co on noble
metal surfaces indicate that charge fluctuations are relevant for these
systems\cite{weiss}.We compared the exact results for the Kondo impurity
with those obtained with an effective model which includes a resonant state
at the Fermi level. Although there are some quantitative differences between
the two models, the exact results are qualitatively reproduced by the simple
one-body approach. We have also calculated the exact spin-spin correlations
in small systems. The results show the existence of magnetic mirages: at the
empty focus conduction electron spins are fluctuating coherently with the
impurity spin. These magnetic mirages are also obtained for non-Kondo
magnetic impurities which couple ferromagnetically with conduction electron
spins. For this case the magnetic mirage is more pronounced than for the
Kondo impurity.

The results obtained here for a sphere can be extended to other systems with
focusing properties such as quantum corrals. In particular, in addition to a
mirage in the localized density of states, a magnetic mirage should be
observed in such systems in the presence of a Kondo (or $sd$) impurity.
Results for the elliptical corral will be presented elsewhere. \newline

This work was done under PICT grant N02151 (ANPCYT). K.H. and C. B. are a
fellows of the Consejo Nacional de Investigaciones Cient\'{\i }ficas y
T\'{e}cnicas (CONICET).


\begin{references}
\bibitem{Mano}  H. C. Manoharan, C. P. Lutz, and D. M. Eigler, Nature {\bf %
403} 512 (2000)

\bibitem{Co}  J. Li, W. D. Schneider, R. Berndt and B. Delley, Phys. Rev.
Lett. {\bf 80}, 2893 (1998)

\bibitem{Cu}  V. Madhavan, W. Chen, T. Jamneala, M. F. Crommie and N. S.
Wingreen, Science {\bf 280}, 567 (1998)

\bibitem{Ag}  O. Agam and A. Schiller, cond-mat/0006443

\bibitem{BW}  M. Weissmann and H. Bonadeo, cond-mat/0007485

\bibitem{Esp}  D. Porras, J. Fern\'{a}ndez-Rossier and C. Tejedor,
cond-mat/0007445

\bibitem{PS}  G. F. Fiete et al, cond-mat/0008170

\bibitem{alig}  A. A. Aligia, cond-mat/0101082

\bibitem{Lancz}  E. Dagotto and A. Moreo, Phys. Rev. D {\bf 31}, 865 (1985)

\bibitem{dyn}  E. R. Gagliano and C. A. Balseiro, Phys. Rev. Lett, {\bf 59},
2999 (1987)

\bibitem{And}  P. W. Anderson, Phys. Rev. {\bf 124}, 41 (1961)

\bibitem{SW}  J. R. Schrieffer and P. A. Wolff, Phys. Rev. {\bf 144}, 491
(1966)

\bibitem{weiss}  M. Weissmann and A. Saul, Phys. Rev. B {\bf 59}, 8405 (1999)
\end{references}
\end{document}